\documentclass[11pt]{article}
\usepackage{amsmath,amssymb,color,graphics,epsfig,cite}
\usepackage{multirow}
\usepackage{ulem}

\usepackage{amsfonts}

\newcommand{\hoch}[1]{$\, ^{#1}$}

\makeatletter
\@addtoreset{equation}{section}
\makeatother

\newcommand{\be}{\begin{equation}}
\newcommand{\ee}{\end{equation}}
\newcommand{\bea}{\setlength\arraycolsep{2pt} \begin{eqnarray}}
\newcommand{\eea}{\end{eqnarray}}
\newcommand{\nn}{\nonumber}

\def\ft#1#2{{\textstyle{\frac{\scriptstyle #1}{\scriptstyle #2} } }}
\def\fft#1#2{{\frac{#1}{#2}}}

\def\0{{\sst{(0)}}}
\def\1{{\sst{(1)}}}
\def\2{{\sst{(2)}}}
\def\3{{\sst{(3)}}}
\def\4{{\sst{(4)}}}
\def\5{{\sst{(5)}}}
\def\6{{\sst{(6)}}}
\def\7{{\sst{(7)}}}
\def\8{{\sst{(8)}}}
\def\sst#1{{\scriptscriptstyle #1}}

\begin{document}

\begin{flushright}
\hfill{}

\end{flushright}

\begin{center}
{\Large {\bf G\"odel Metrics with Chronology Protection\\ in Horndeski Gravities}}

\vspace{10pt}
Wei-Jian Geng\hoch{1},  Shou-Long Li\hoch{2},  H. L\"u\hoch{3*} and Hao Wei\hoch{2}

\vspace{10pt}

\hoch{1}{\it Department of Physics, Beijing Normal University, Beijing 100875, China}

\vspace{10pt}

\hoch{2}{\it School of Physics,
Beijing Institute of Technology, Beijing 100081, China}

\vspace{10pt}
\hoch{3}{\it Department of Physics, Tianjin University, Tianjin 300350, China}

\vspace{40pt}

\underline{ABSTRACT}

\end{center}

G\"odel universe, one of the most interesting exact solutions predicted by General Relativity, describes a homogeneous rotating universe containing naked closed time-like curves (CTCs). It was shown that such CTCs are the consequence of the null energy condition in General Relativity.  In this paper, we show that the G\"odel-type metrics with chronology protection can emerge in Einstein-Horndeski gravity.  We construct such exact solutions also in Einstein-Horndeski-Maxwell and Einstein-Horndeski-Proca theories.

\vfill {\footnotesize gengwj@mail.bnu.edu.cn \ \ \ sllee\_phys@bit.edu.cn \ \ \ \hoch{*}mrhonglu@gmail.com
\ \ \ haowei@bit.edu.cn}

\thispagestyle{empty}

\pagebreak





\section{Introduction}\label{sec1}

G\"odel metric \cite{Godel:1949ga}, constructed in 1949, is one of the most interesting exact solutions predicted by General Relativity~(GR). The original construction by G\"odel requires a fine-tuning balance between a negative cosmological constant and the matter density of some uniform homogeneous pressureless perfect fluids or dusts.  The G\"odel universe is a direct product of a three-dimensional homogeneous rotating spacetime with a real line. This spacetime metric exhibits several peculiar features, such as the presence of naked closed time-like curves~(CTCs) and the absence of globally spatial-like Cauchy surface.

The G\"odel metric can be generalized by introducing a constant $\alpha$, and the resulting metrics are referred as the G\"odel type. Naked CTCs are present as long as $0<\alpha<1$, with $\alpha=1/2$ corresponding to the original G\"odel metric. (See section 2 for details.)  In this paper, we consider only these four-dimensional {\it G\"odel-type metrics}, and we believe that it is appropriate to call all these metrics simply as {\it G\"odel metrics}. There have been continuing efforts in the GR community to construct and study the G\"odel metrics, see, for example,
~\cite{Banerjee,Bampi, Raychaudhuri:1980fd,Reboucas:1982hn,Reboucas:1986tz, Barrow:1998wa,Kanti:1999am,Israel:2003cx,Gurses:2003cm, Gurses:2005ru, Gleiser:2005nw,Reboucas:2009yw, Santos:2010tw,Ahmedov:2010fz,Furtado:2011tc,Liu:2012kka,Fonseca-Neto:2013rna,Agudelo:2016pic,
Porfirio:2016nzr,Li:2016nnn,Dadhich:2017zdi, Gama:2017eip} and references therein.  G\"odel metrics can also be embedded in strings and M-theory \cite{Israel:2003cx,Li:2016nnn}. (In literature, the G\"odel metric was generalized to include two parameters; however, as was demonstrated in \cite{Li:2016nnn}, these metrics are related to the metric of the single-parameter $\alpha$ by coordinate transformations.)

The existence of naked CTCs in spacetime violates the chronology protection conjecture proposed by Hawking~\cite{Hawking:1991nk}. Since G\"odel metrics can be supported by fundamental matter fields such as Maxwell and axion fields, microscopic G\"odel metrics present challenges to the chronological protection.  G\"odel metrics are absent from CTCs when the parameter $\alpha\ge 1$.  What is intriguing is that with the framework of Einstein gravity with minimally-coupled matter, the null energy condition implies that $0< \alpha\le 1$ \cite{Li:2016nnn}.  Thus naked CTCs are unavoidable for G\"odel metrics in Einstein gravity, unless $\alpha=1$, which corresponds to simply a direct product of a real line and anti-de Sitter spacetime (AdS) in three dimensions. The general $\alpha>1$ metrics with chronology protection requires modified gravities beyond Einstein gravity \cite{Barrow:1998wa,Reboucas:2009yw,Ahmedov:2010fz,Santos:2010tw,Furtado:2011tc,
Fonseca-Neto:2013rna,Agudelo:2016pic,Porfirio:2016nzr}.  In particular,  in the low-energy effective theory of strings with higher-derivative corrections, G\"odel metrics with $\alpha >1$ was constructed in \cite{Barrow:1998wa}. However, this theory, when treated on its own, involves ghost modes. It is thus of interest to look for ghost-free theories beyond Einstein gravity that support G\"odel metrics with $\alpha>1$ so that the spacetime is absent from CTCs.

In this paper, we construct G\"odel metrics in Einstein-Horndeski gravity. The Horndeski terms \cite{Horndeski:1974wa} involve a non-minimally coupled axionic scalar that enters the Lagrangian only through a covariant derivative.  The theory is analogous to Gauss-Bonnet gravity where the linearized equations of motion are of two derivatives.  Horndeski gravity has been deeply investigated in the context of cosmology, (see e.g.~\cite{Amendola:1993uh,Germani:2010gm};) and also in the context of black holes \cite{ac,aco,Rinaldi:2012vy,Babichev:2013cya},  black hole thermodynamics \cite{Feng:2015oea, Feng:2015wvb}, and holographic properties \cite{Jiang:2017imk,Baggioli:2017ojd,Caceres:2017lbr,Liu:2017kml}.  Recently new black holes were constructed that violate \cite{Feng:2017jub} the conjecture of the reverse isoperimetric inequality proposed in \cite{Cvetic:2010jb}.

In the embedding of G\"odel metrics in string and M-theory, an axion carrying a proper magnetic axion charge plays an important role \cite{Li:2016nnn}.  It is thus natural to consider Horndeski gravity where an axion is a necessary component in the theory.  Furthermore, as a modified theory to Einstein gravity, its null energy condition is modified, and hence Horndeski gravity may restore the chronological protection, and we find that it is indeed the case. Chronological protection in general classes of Horndeski gravities were also discussed in \cite{Burrage:2011cr}.

The paper is organized as follows. In section \ref{sec2}, we consider G\"odel metrics with general $\alpha$ parameter and review that $\alpha\ge 1$ metrics are chronologically protected.  We then demonstrate that such metrics can arise from Einstein-Horndeski gravity where the Horndeski axion carries a magnetic charge. In section \ref{sec3}, we obtain the G\"odel metrics in Einstein-Horndeski-Maxwell and Einstein-Horndeski-Proca theories. We conclude the paper in section \ref{sec4}.

\section{G\"odel metrics in Horndeski theory}\label{sec2}

In this paper, we consider a class of metrics in four dimensions that take the form
\be
ds^2 = \ell^2 \left[-(dt+ r d\phi)^2 +\alpha r^2 d\phi^2 +\frac{dr^2}{r^2} + dz^2 \right]  \,,  \label{Godel}
\ee
where $\ell$ and $\alpha>0$ are constants. The metric is homogeneous and a direct product of a real line $\mathbb R$ and a three-dimensional rotating spacetime which was called G$_\alpha$ in \cite{Li:2016nnn}.
The G$_\alpha$ associates with the three dimensional metric of $(t, \phi, r)$ and the $\mathbb R$ associates with the coordinate $z$. When $\alpha= 1/2$, the original G\"odel metric is recovered, corresponding to G$_{1/2} \times \mathbb R$.  The solution can be constructed by fine-tuning a negative cosmological constant against homogeneous pressureless perfect fluids \cite{Godel:1949ga}. When $\alpha=1$, the G$_\alpha$ part of the metric is locally AdS$_3$, i.e. G$_1=$ AdS$_3$. In this paper, we find it appropriate to refer all the G$_\alpha\times {\mathbb R}$ metrics (\ref{Godel}) as G\"odel metrics.

    For $\alpha\ge 1$, $t$ is the globally-defined time coordinate. This is no longer true when
$\alpha<1$, which is indicative of the existence of possible CTCs.  To see this explicitly, one can make a coordinate transformation \cite{Li:2016nnn}
\bea
&&r=\cosh \hat r + \cos\hat \phi\, \sinh \hat r\,,\qquad
r\phi =\fft1{\sqrt\alpha} \sin\hat\phi\,\sinh\hat r\,,\nn\\
&& \tan\big(\ft12\hat \phi + \ft12\sqrt\alpha\, (t - \hat t)\big) = e^{-\hat r}
\tan(\ft12 \hat \phi)\,,\label{trans}
\eea
the G\"odel metrics (\ref{Godel}) become
\be
ds^2 = \ell^2\Big[-  \left(d\hat t + \ft{2}{\sqrt{\alpha}}\,\sinh^2(\ft12 \hat r)\, d\hat \phi\right)^2 + \sinh^2\hat r\, d\hat \phi^2 + d\hat r^2 + dz^2\Big]\,.\label{godel2}
\ee
Absence of a conic singularity at $\hat r=0$ requires that $\hat\phi$ be periodic and the period be $\Delta\hat\phi=2\pi$. It follows that for $\alpha <1$ the spacetime develops negative $g_{\hat\phi\hat\phi}$ for sufficiently large $r$, indicating naked CTCs.  Such CTCs are absent for G\"odel metrics with $\alpha\ge1$. In the previous related works~\cite{Li:2016nnn}, we found within the framework of Einstein gravity, the null energy condition requires that $\alpha\leq1$, and hence the G\"odel metrics in Einstein gravity necessary have naked CTCs.

It should be emphasized that the three-dimensional metric $G_\alpha$ is completely specified by its Ricci tensor.  Thus, as was remarked in the introduction, the two-parameter G\"odel metrics in literature can be all derived by some coordinate transformations from our $G_\alpha$ metric.

In this section, we show that solutions with $\alpha>1$ can emerge in Einstein-Horndeski theory. Horndeski terms are a class of higher derivative polynomials constructed from Riemann tensors and axionic scalars \cite{Horndeski:1974wa}. The action of Einstein-Horndeski gravity at the lowest-order in four dimensions is given by
\be
{\cal S} = \int \sqrt{-g} \, L \, d^4 x \,, \quad L =  \kappa (R-2\Lambda) -\frac12 \left(\beta \, g_{\mu\nu}- \gamma \, G_{\mu\nu} \right) \nabla^\mu\chi\nabla^\nu\chi +L^{\rm mat}\,, \label{Hornd}
\ee
where $\kappa$, $\beta$ and $\gamma$ are coupling constants, $G_{\mu\nu} = R_{\mu\nu} -\frac{1}{2} R g_{\mu\nu}$ is the Einstein tensor, $\chi$ is axionic scalar field, and $L^{\rm mat}$ is the Lagrangian of matter. When $\gamma=0, \beta=0$ and $\kappa=1$, the theory reduces to Einstein theory with a cosmological constant. When the axion $\chi$ is constant, the Einstein theory is also recovered. The explicit Einstein and axion equations can be found in literature, see, e.g.~\cite{aco,Feng:2015oea}.  For simplicity, we take $\kappa=1$ throughout the paper.

Since the G\"odel metric (\ref{Godel}) is homogeneous, the coefficient $m_0$ in the kinetic term for the axion, namely $K=\ft12 m_0^2\, \dot \chi^2$, is constant, and it must be non-negative.  In other words, the ghost-free condition requires that
\be
m_0^2=- \left(\beta \, \eta^{00} - \gamma \, G^{00}\right) =  \frac{4\alpha \beta \ell^2 + (3 - 4 \alpha) \gamma}{4 \alpha \ell^2} \ge 0 \,. \label{energy}
\ee

For the general G\"odel metric (\ref{Godel}) , we follow the analogous construction of \cite{Li:2016nnn} and take the axion to be magnetic:
\be
\chi = k z \,.
\ee
For pure Einstein-Horndeski theory without matter, we obtain the solution
\be
\gamma=\frac{4\alpha\beta \ell^2}{4\alpha-1} \,,\quad k=\sqrt{\frac{4\alpha-1}{\alpha\beta}} \,,\quad \Lambda=-\frac{4\alpha-1}{4\alpha \ell^2} \,. \label{solution}
\ee
Substituting the solution into the ghost-free condition (\ref{energy}), we find
\be
m_0^2=\frac{2\beta}{4\alpha-1} >0 \,.
\ee
Together with the reality condition for constant $k$, we find two branches of solutions
\be
\begin{cases}
	\alpha>\frac14 \\
	\beta>0
\end{cases}\qquad\qquad
\textup{or} \qquad\qquad\quad
\begin{cases}
	0<\alpha<\frac14 \\
	\beta<0
\end{cases} \,.
\ee
Thus we see that the original ($\alpha=1/2$) G\"odel metric can emerge in Horndeski gravity; furthermore, in addition to the usual G\"odel metrics with $\alpha<1$, metrics with $\alpha>1$ that maintain the chronological protection can also emerge.

\section{G\"odel metrics in Horndeski theory with matter}\label{sec3}

In this section, we generalized the G\"odel metrics in Horndeski gravity by introducing matter fields, such as Maxwell and Proca fields.  We also restrict the constant $\beta$ to be positive only.  In some cases, we can set $\beta=1$ without loss of generality.

\subsection{Maxwell field}
The Lagrangian for the Maxwell field is given by
\be
L^{\rm mat} = -\ft14 F^2\,,\qquad F=dA\,.
\ee
Assuming that the coordinate $z$ is periodic, one can take the following ansatz \cite{Reboucas:1982hn}:
\be
A=q \sin(w z) (dt +rd\phi)  \,,
\ee
where $w$ is a constant, inverse to the period of coordinate $z$. The solution is best described as G$_{\alpha}\times S^1$ rather than G$_{\alpha}\times \mathbb{R}$. For the metric~(\ref{Godel}), the solutions are given by
\bea
&&w =\frac{1}{\sqrt{\alpha}} \,, \qquad \Lambda =\frac{(1-2\alpha)(1-4\alpha)\gamma-8\alpha^2\beta \ell^2}{8 \alpha\, \ell^2 \big((1-2\alpha) \gamma +2\alpha \beta \ell^2\big)}  \,, \\
&&q^2 = \frac{\ell^2 (\alpha-1) \big((4\alpha-1)\gamma -4\alpha \beta \ell^2\big)}{(1-2\alpha)\gamma+2\alpha\beta \ell^2}  \,,  \qquad k =\frac{\sqrt{2} \,\ell}{\sqrt{(1-2\alpha)\gamma+2\alpha\beta \ell^2}}  \,.
\eea
Solutions with $\gamma=0$ were constructed in \cite{Reboucas:1982hn}. The reality conditions of $(q,k)$ imply that
\be
(\alpha-1)\big((4 \alpha-1)\gamma - 4 \alpha \beta \ell^2\big)\ge 0\,,\qquad
(2 \alpha-1)\gamma - 2 \alpha \beta \ell^2<0\,.
\ee
When $\alpha=1$, Maxwell field vanishes, reverting back to a special case discussed in the previous section. For $\alpha>1$, we find that $\gamma$ must lie within the range:
\be
\alpha>1:\qquad\qquad \frac{4\alpha\beta \ell^2}{4\alpha-1}\le \gamma< \frac{2\alpha\beta \ell^2}{2\alpha-1} \,.
\ee
Substituting the $\gamma$ range into (\ref{energy}), we find that the ghost free condition is indeed satisfied, namely
\be
0<\frac{\beta}{2(2\alpha-1)}\le m_0^2< \frac{2\beta}{4\alpha-1} \,.
\ee

G\"odel metrics with $\alpha <1$ can also arise.  In this case, the reality conditions imply that
\be
(4 \alpha-1)\gamma - 4 \alpha \beta \ell^2\le 0\,,\qquad (2 \alpha-1)\gamma - 2 \alpha \beta \ell^2<0\,.
\ee
Together with the ghost-free condition (\ref{energy}), we find that $\alpha<1$ solutions are also possible in Einstein-Horndeski-Maxwell gravity and the corresponding $\gamma$ range is given by
\bea
 \begin{tabular}{|c|c|c|c|}
   \hline $0<\alpha \le  \fft14$ &$\fft14<\alpha< \fft34$
   & $\fft34\le \alpha <1$ \\
   \hline $\gamma\ge - \fft{4\alpha\beta\ell^2}{3-4\alpha}$ &$-\fft{4\alpha\beta\ell^2}{3-4\alpha}\le \gamma \le \fft{4\alpha\beta\ell^2}{4\alpha-1}$   & $\gamma\le
   \fft{4\alpha\beta\ell^2}{4\alpha-3}$\\
   \hline
 \end{tabular}\,.
\eea

\subsection{Including $\chi \, F \wedge F$}
We can further add a topological term so that the matter content becomes,
\be
L^{mat} = -\ft14 F^2 + \ft18 \chi \epsilon ^{\mu\nu\rho\sigma}F_{\mu\nu}F_{\rho\sigma} \,.
\ee
The coordinate $z$ is now treated as a real line, instead as a circle.  Making an ansatz for the Maxwell potential \cite{Li:2016nnn}
\be
A=q rd\phi  \,,
\ee
we find that the solutions are given by
\bea
&&k =\frac{1}{\sqrt{\alpha}} \,, \quad \Lambda = -\fft{8\alpha^2\ell^2+(2\alpha-1)\gamma}{16\alpha^2 \ell^4} \,, \cr
&&q^2 = \frac{(1-\alpha) \left(4 \alpha  \ell^2-\gamma \right)}{2 \alpha } \,, \quad \beta = 1+\frac{(2 \alpha-1)\gamma}{2 \alpha  \ell^2} \,.
\eea
Solutions with $\gamma=0$ were constructed in \cite{Li:2016nnn}. We require that $\beta>0$, and together with ghost-free condition (\ref{energy}) and reality condition, we find that the results are:
\bea
 \begin{tabular}{|c|c|c|c|}
   \hline $0<\alpha \le \fft14$ &$\fft14 <\alpha <\fft34$ & $\fft34\le\alpha\le 1$
   &$\alpha >1$\\ \hline
   $-4\alpha\ell^2\le \gamma < \fft{2\alpha\ell^2}{1-2\alpha}$ &$|\gamma|\le 4 \alpha \ell^2 $ & $-\frac{2 \alpha \ell^2}{2\alpha-1} < \gamma \le 4 \alpha \ell^2$ & $\gamma\ge 4\alpha \ell^2$\\
   \hline
 \end{tabular}\,.
\eea

\subsection{Proca field}

We now consider the Proca field, and the Lagrangian is given by
\be
L^{\rm mat} = - \ft14 F^2 - \ft12 m^2 A^2\,,
\ee
where $F=dA$.  Taking the ansatz \cite{Li:2016nnn}
\be
A=q (dt +rd\phi)  \,,
\ee
we find general G\"odel solutions with
\begin{align}
m &=\frac{1}{\sqrt{\alpha} \, \ell}  \,, \quad \Lambda =\frac{(1-2\alpha)(1-4\alpha) \gamma +4 (1-3\alpha) \alpha \beta \ell^2}{4\alpha \ell^2 (\gamma-3\alpha\gamma +4\alpha \beta \ell^2)}   \,, \\
q^2 &=\frac{\ell^2 (\alpha-1) (\gamma(4\alpha-1) -4\alpha \beta \ell^2)}{\gamma(1-3\alpha)+4\alpha \beta \ell^2}   \,, \quad k =2 \,\ell \sqrt{\frac{\alpha}{\gamma(1-3\alpha)+4\alpha\beta \ell^2}}  \,.
\end{align}
Solutions with $\gamma=0$ were constructed in \cite{Li:2016nnn}. The reality condition, together with the ghost-free condition (\ref{energy}), restrict the parameter regions.  For our choice of $\beta>0$, we find that the chronology-protected $(\alpha >1)$ G\"odel metrics exist in Einstein-Horndeski-Proca theory, and the Horndeski coupling constant $\gamma$ lies in the following regions:
\bea
 \begin{tabular}{|c|c|}
   \hline $1<\alpha \le 2$ &$\alpha >2$ \\
   \hline $\frac{4 \alpha \beta \ell^2}{4 \alpha -1} < \gamma < \frac{4 \alpha \beta \ell^2}{3\alpha-1} $ &$\frac{4 \alpha \beta \ell^2}{4 \alpha -1} < \gamma \le \frac{4 \alpha \beta \ell^2}{4\alpha-3}$  \\
   \hline
 \end{tabular}\,.
\eea
G\"odel metrics with $\alpha\le 1$ can also arise and the results for the constant $\gamma$ range is identical to those in the Einstein-Maxwell theory discussed earlier.

\section{Conclusions}\label{sec4}

In this paper, we considered a class of G\"odel metrics (\ref{Godel}) with a free parameter $\alpha$.
In Einstein gravity, the null energy condition imposes that $\alpha\le 1$, with $\alpha=1$ corresponding to AdS$_3\times \mathbb R$.  Since G\"odel metrics with $\alpha <1$ have naked CTCs, the chronology in these universes are not protected.

We constructed G\"odel metrics in Einstein-Horndeski theories, with or without additional matter that includes Maxwell and Proca fields.  We find that G\"odel metrics with $\alpha>1$ that evade CTCs can also emerge, and consequently the corresponding chronology is protected.  We determine the range of the Horndeski coupling constant $\gamma$ so that the theories that admit the G\"odel metrics are absent from ghost excitations.  The full stability of these metrics however requires further investigation.

\section*{Acknowledgement}

W.-J.G.~and H.L.~are supported in part by NSFC grants No.~11475024, No.~11175269 and No.~11235003.
S.-L.L. is supported in part by Graduate Technological Innovation Project of
Beijing Institute of Technology. H.W. is supported in part by NSFC grants No.~11575022
and No.~11175016.


\begin{thebibliography}{99}

\bibitem{Godel:1949ga}
  K.~G\"odel,
{\it An example of a new type of cosmological solutions of Einstein's field equations of graviation,}
  Rev.\ Mod.\ Phys.\  {\bf 21}, 447 (1949).
  doi:10.1103/RevModPhys.21.447

\bibitem{Banerjee}
  A.~Banerjee and S.~Banerji,
{\it Stationary distributions of dust and electromagnetic fields in general relativity,}
  J.\ Phys.\ A {\bf 1}, 188 (1968).

\bibitem{Bampi}
F.~Bampi and C.~Zordan,
{\it A note on G\"odel's metric},
Gen.\ Rel.\ Grav., {\bf 9}, 393 (1978).

\bibitem{Raychaudhuri:1980fd}
  A.K.~Raychaudhuri and S.N.~Guha Thakurta,
  {\it Homogeneous space-times of the G\"odel type,}
  Phys.\ Rev.\ D {\bf 22}, 802 (1980).
  doi:10.1103/PhysRevD.22.802

\bibitem{Reboucas:1982hn}
  M.J.~Reboucas and J.~Tiomno,
  {\it On the homogeneity of Riemannian space-times of G\"odel type,}
  Phys.\ Rev.\ D {\bf 28}, 1251 (1983).
  doi:10.1103/PhysRevD.28.1251

\bibitem{Reboucas:1986tz}
  M.J.~Reboucas, J.E.~Aman and A.F.F.~Teixeira,
  {\it A note on G\"odel type space-times,}
  J.\ Math.\ Phys.\  {\bf 27}, 1370 (1986).
  doi:10.1063/1.527093

\bibitem{Barrow:1998wa}
  J.D.~Barrow and M.P.~Dabrowski,
{\it G\"odel universes in string theory,}
  Phys.\ Rev.\ D {\bf 58}, 103502 (1998)
  doi:10.1103/PhysRevD.58.103502
  [gr-qc/9803048].

\bibitem{Kanti:1999am}
P.~Kanti and C.E.~Vayonakis,
{\it G\"odel type universes in string inspired charged gravity,}
Phys.\ Rev.\ D {\bf 60}, 103519 (1999)
doi:10.1103/PhysRevD.60.103519
[gr-qc/9905032].

\bibitem{Israel:2003cx}
  D.~Israel,
{\it Quantization of heterotic strings in a Godel/anti-de Sitter space-time
and chronology protection,}
  JHEP {\bf 0401}, 042 (2004)
  doi:10.1088/1126-6708/2004/01/042
  [hep-th/0310158].

\bibitem{Gurses:2003cm}
  M.~Gurses, A.~Karasu and O.~Sarioglu,
  {\it Godel type of metrics in various dimensions,}
  Class.\ Quant.\ Grav.\  {\bf 22}, 1527 (2005)
  doi:10.1088/0264-9381/22/9/003
  [hep-th/0312290].

\bibitem{Gurses:2005ru}
  M.~Gurses and O.~Sarioglu,
  {\it Godel-type metrics in various dimensions II: Inclusion of a dilaton field,}
  Class.\ Quant.\ Grav.\  {\bf 22}, 4699 (2005)
  doi:10.1088/0264-9381/22/22/004
  [hep-th/0505268].

\bibitem{Gleiser:2005nw}
  R.J.~Gleiser, M.~Gurses, A.~Karasu and O.~Sarioglu,
  {\it Closed timelike curves and geodesics of Godel-type metrics,}
  Class.\ Quant.\ Grav.\  {\bf 23}, 2653 (2006)
  doi:10.1088/0264-9381/23/7/025
  [gr-qc/0512037].

\bibitem{Reboucas:2009yw}
  M.J.~Reboucas and J.~Santos,
{\it G\"odel-type universes in $f(R)$ gravity,}
  Phys.\ Rev.\ D {\bf 80}, 063009 (2009)
  doi:10.1103/PhysRevD.80.063009
  [arXiv:0906.5354 [astro-ph.CO]].

\bibitem{Ahmedov:2010fz}
  H.~Ahmedov and A.N.~Aliev,
  {\it Black string and G\"odel type solutions of Chern-Simons modified gravity,}
  Phys.\ Rev.\ D {\bf 82}, 024043 (2010)
  doi:10.1103/PhysRevD.82.024043
  [arXiv:1003.6017 [hep-th]].

\bibitem{Santos:2010tw}
  J.~Santos, M.J.~Reboucas and T.B.R.F.~Oliveira,
  {\it G\"odel-type universes in Palatini $f(R)$ gravity,}
  Phys.\ Rev.\ D {\bf 81}, 123017 (2010)
  doi:10.1103/PhysRevD.81.123017
  [arXiv:1004.2501 [astro-ph.CO]].

\bibitem{Furtado:2011tc}
  C.~Furtado, J.R.~Nascimento, A.Y.~Petrov and A.F.~Santos,
  {\it Horava-Lifshitz gravity and G\"odel universe,}
  Phys.\ Rev.\ D {\bf 84}, 047702 (2011)
  Erratum: [Phys.\ Rev.\ D {\bf 84}, 069904 (2011)]
  doi:10.1103/PhysRevD.84.047702, 10.1103/PhysRevD.84.069904
  [arXiv:1106.4003 [hep-th]].

\bibitem{Liu:2012kka}
  D.~Liu, P.~Wu and H.~Yu,
  {\it G\"{o}del-type universes in $f(T)$ gravity,}
  Int.\ J.\ Mod.\ Phys.\ D {\bf 21}, 1250074 (2012)
  doi:10.1142/S0218271812500745
  [arXiv:1203.2016 [gr-qc]].

\bibitem{Fonseca-Neto:2013rna}
  J.B.~Fonseca-Neto, A.Y.~Petrov and M.J.~Reboucas,
  {\it G\"odel-type universes and chronology protection in Horava-Lifshitz gravity,}
  Phys.\ Lett.\ B {\bf 725}, 412 (2013)
  doi:10.1016/ j.physletb.2013.07.018
  [arXiv:1304.4675 [astro-ph.CO]].

\bibitem{Agudelo:2016pic}
  J.A.~Agudelo, J.R.~Nascimento, A.Y.~Petrov, P.J.~Porf\'irio and A.F.~Santos,
{\it G\"odel and G\"odel-type universes in Brans-Dicke theory,}
  Phys.\ Lett.\ B {\bf 762}, 96 (2016)
  [arXiv: 1603.07582 [hep-th]].

\bibitem{Porfirio:2016nzr}
  P.J.~Porfirio, J.B.~Fonseca-Neto, J.R.~Nascimento, A.Y.~Petrov, J.~Ricardo and A.F.~Santos,
{\it Chern-Simons modified gravity and closed timelike curves,}
  Phys.\ Rev.\ D {\bf 94}, no. 4, 044044 (2016)
  doi:10.1103/PhysRevD.94.044044
  [arXiv:1606.00743 [hep-th]].

\bibitem{Li:2016nnn}
  S.L.~Li, X.H.~Feng, H.~Wei and H.~L\"u,
  {\it G\"odel universe from string theory,}
  Eur.\ Phys.\ J.\ C {\bf 77}, no. 5, 289 (2017)
  doi:10.1140/epjc/s10052-017-4856-z
  [arXiv:1612.02069 [hep-th]].

\bibitem{Dadhich:2017zdi}
  N.~Dadhich, A.~Molina and J.M.~Pons,
  {\it Generalized G\"odel universes in higher dimensions and pure Lovelock gravity,}
  Phys.\ Rev.\ D {\bf 96}, no. 8, 084058 (2017)
  doi:10.1103/PhysRevD.96.084058
  [arXiv:1703.05663 [gr-qc]].

\bibitem{Gama:2017eip}
  F.S.~Gama, J.R.~Nascimento, A.Y.~Petrov, P.J.~Porfirio and A.F.~Santos,
  {\it G\"odel-type solutions within the f(R,Q) gravity,}
  Phys.\ Rev.\ D {\bf 96}, no. 6, 064020 (2017)
  doi:10.1103/PhysRevD.96.064020
  [arXiv:1707.03440 [hep-th]].

\bibitem{Hawking:1991nk}
  S.W.~Hawking,
  {\it The chronology protection conjecture,}
  Phys.\ Rev.\ D {\bf 46}, 603 (1992).
  doi:10.1103/PhysRevD.46.603


\bibitem{Horndeski:1974wa}
  G.W.~Horndeski,
  {\it Second-order scalar-tensor field equations in a four-dimensional space,}
  Int.\ J.\ Theor.\ Phys.\  {\bf 10}, 363 (1974).
  doi:10.1007/BF01807638


\bibitem{Amendola:1993uh}
L.~Amendola,
{\it Cosmology with nonminimal derivative couplings,}
Phys.\ Lett.\ B {\bf 301}, 175 (1993)
doi:10.1016/0370-2693(93)90685-B
[gr-qc/9302010].

\bibitem{Germani:2010gm}
C.~Germani and A.~Kehagias,
{\it New model of inflation with non-minimal derivative coupling of Standard Model Higgs boson to gravity,}
Phys.\ Rev.\ Lett.\  {\bf 105}, 011302 (2010)
doi:10.1103/PhysRevLett.105.011302
[arXiv:1003.2635 [hep-ph]].

\bibitem{ac}
  A.~Cisterna and C.~Erices,
{\it Asymptotically locally AdS and flat black holes in the presence of an electric field in the Horndeski scenario,}
  Phys.\ Rev.\ D {\bf 89}, 084038 (2014)
  doi:10.1103/PhysRevD.89.084038
  [arXiv:1401.4479 [gr-qc]].

\bibitem{aco}
  A.~Anabalon, A.~Cisterna and J.~Oliva,
{\it Asymptotically locally AdS and flat black holes in Horndeski theory,}
  Phys.\ Rev.\ D {\bf 89}, 084050 (2014) doi:10.1103/PhysRevD.89.084050
arXiv:1312.3597 [gr-qc].

\bibitem{Rinaldi:2012vy}
  M.~Rinaldi,
{\it Black holes with non-minimal derivative coupling,}
  Phys.\ Rev.\ D {\bf 86}, 084048 (2012)
doi:10.1103/PhysRevD.86.084048
arXiv:1208.0103 [gr-qc].

\bibitem{Babichev:2013cya}
  E.~Babichev and C.~Charmousis,
{\it Dressing a black hole with a time-dependent Galileon,}
  JHEP {\bf 1408}, 106 (2014)
doi:10.1007/JHEP08(2014)106
arXiv:1312.3204 [gr-qc].


\bibitem{Feng:2015oea}
  X.H.~Feng, H.S.~Liu, H.~L\"u and C.N.~Pope,
  {\it Black hole entropy and viscosity bound in Horndeski gravity,}
  JHEP {\bf 1511}, 176 (2015)
 doi:10.1007/JHEP11(2015)176
  [arXiv:1509.07142 [hep-th]].

\bibitem{Feng:2015wvb}
  X.H.~Feng, H.S.~Liu, H.~L\"u and C.N.~Pope,
  {\it Thermodynamics of charged black holes in Einstein-Horndeski-Maxwell theory,}
  Phys.\ Rev.\ D {\bf 93}, no. 4, 044030 (2016)
  doi:10.1103/PhysRevD.93.044030
  [arXiv:1512.02659 [hep-th]].

\bibitem{Jiang:2017imk}
  W.J.~Jiang, H.S.~Liu, H.~L\"u and C.N.~Pope,
{\it DC conductivities with momentum dissipation in Horndeski theories,}
  JHEP {\bf 1707}, 084 (2017)
  doi:10.1007/JHEP07(2017)084
  [arXiv:1703.00922 [hep-th]].

\bibitem{Baggioli:2017ojd}
  M.~Baggioli and W.J.~Li,
{\it Diffusivities bounds and chaos in holographic Horndeski theories,}
  JHEP {\bf 1707}, 055 (2017)
  doi:10.1007/JHEP07(2017)055
  [arXiv:1705.01766 [hep-th]].

\bibitem{Caceres:2017lbr}
  E.~Caceres, R.~Mohan and P.H.~Nguyen,
{\it On holographic entanglement entropy of Horndeski black holes,}
  JHEP {\bf 1710}, 145 (2017)
  doi:10.1007/JHEP10(2017)145
  [arXiv:1707.06322 [hep-th]].

\bibitem{Liu:2017kml}
  H.S.~Liu, H.~L\"u and C.N.~Pope,
  {\it Holographic heat current as Noether current,}
  JHEP {\bf 1709}, 146 (2017)
  doi:10.1007/JHEP09(2017)146
  [arXiv:1708.02329 [hep-th]].


\bibitem{Feng:2017jub}
  X.H.~Feng, H.S.~Liu, W.T.~Lu and H.~L\"u,
{\it Horndeski gravity and the violation of reverse isoperimetric inequality,}
  Eur.\ Phys.\ J.\ C {\bf 77}, no. 11, 790 (2017)
  doi:10.1140/ epjc/s10052-017-5356-x
  [arXiv:1705.08970 [hep-th]].

\bibitem{Cvetic:2010jb}
  M. Cveti\v c, G.W.~Gibbons, D.~Kubiznak and C.N.~Pope,
  {\it Black hole enthalpy and an entropy inequality for the thermodynamic volume,}
  Phys.\ Rev.\ D {\bf 84}, 024037 (2011)
  doi:10.1103/PhysRevD.84.024037
  [arXiv:1012.2888 [hep-th]].

\bibitem{Burrage:2011cr}
  C.~Burrage, C.~de Rham, L.~Heisenberg and A.J.~Tolley,
{\it Chronology protection in Galileon models and massive gravity,}
  JCAP {\bf 1207}, 004 (2012)
  doi:10.1088/1475-7516/2012/07/004
  [arXiv:1111.5549 [hep-th]].


\end{thebibliography}
\end{document}